\documentclass{emulateapj}
\usepackage{graphicx,amsmath,amssymb}
\usepackage{natbib}
\usepackage{apjfonts}
\bibpunct[, ]{(}{)}{;}{a}{}{,}

\newcommand{\thisgrb}{GRB\,031203}

\newcommand{\sect}{\S\,}

\begin{document}

   \title{A very low luminosity X-ray flash: \emph{XMM-Newton} observations of \thisgrb}

   \author{D.~Watson,\altaffilmark{1} J.~Hjorth,\altaffilmark{1} A.~Levan,\altaffilmark{2} P.~Jakobsson,\altaffilmark{1} P.~T.~O'Brien,\altaffilmark{2} J.~P.~Osborne,\altaffilmark{2} K.~Pedersen,\altaffilmark{1} J.~N.~Reeves,\altaffilmark{3,4} J.~A.~Tedds,\altaffilmark{2} S.~A.~Vaughan,\altaffilmark{2} M.~J.~Ward,\altaffilmark{2} and R. Willingale\altaffilmark{2}
           }
   \altaffiltext{1}{Niels Bohr Institute, Astronomical Observatory, University of Copenhagen, Juliane-Maries Vej 30, DK-2100 Copenhagen \O, Denmark; darach, jens, pallja, kp @astro.ku.dk}
   \altaffiltext{2}{X-Ray Astronomy Group, Department of Physics and Astronomy, Leicester University, Leicester LE1 7RH, UK; anl, pto, julo, jat, sav2, mjw, rw @star.le.ac.uk}
   \altaffiltext{3}{Laboratory for High Energy Astrophysics, Code 662, NASA Goddard Space Flight Center, Greenbelt, MD 20771, USA; jnr@milkyway.gsfc.nasa.gov}
   \altaffiltext{4}{Universities Space Research Association}

   \begin{abstract}
     \thisgrb\ was observed by \emph{XMM-Newton} twice, first with an
     observation beginning 6\,hours after the burst, and again after 3\,days.
     The afterglow had average 0.2--10.0\,keV fluxes for the first and
     second observations of $4.2\pm0.1\times10^{-13}$ and
     $1.8\pm0.1\times10^{-13}$\,erg\,cm$^{-2}$\,s$^{-1}$ respectively, decaying
     very slowly according to a power-law with an index of $-0.55\pm0.05$. 
     The prompt soft X-ray flux, inferred from a detection of the dust echo
     of the prompt emission, strongly implies that this burst is very soft
     and should be classified as an X-ray flash (XRF) and further, implies a
     steep temporal slope ($\lesssim-1.7$) between the prompt and afterglow
     phases or in the early afterglow, very different from the later
     afterglow decay slope.  A power-law ($\Gamma=1.90\pm0.05$) with
     absorption at a level consistent with the Galactic foreground
     absorption fits the afterglow spectrum well. A bright, low-redshift
     ($z=0.105$) galaxy lies within 0.5\arcsec\ of the X-ray position and is
     likely to be the GRB host.  At this redshift,
     \thisgrb\ is the closest GRB or XRF known after GRB\,980425.  It has a
     very low equivalent isotropic $\gamma$-ray energy in the burst
     ($\sim3\times10^{49}$\,erg) and X-ray luminosity in the afterglow
     ($9\times10^{42}$\,erg\,s$^{-1}$ at 10\,hours), 3--4 orders of
     magnitude less than typical bursts, though higher than either the faint
     XRF\,020903 or GRB\,980425.  The rapid initial decline and subsequent
     very slow fading of the X-ray afterglow is also similar to that
     observed in GRB\,980425, indicating that \thisgrb\ may be
     representative of low luminosity bursts.
   \end{abstract}

   \maketitle

%
%
\section{Introduction\label{introduction}}

A class of very soft burst, very similar to GRBs, has been identified and
are referred to as X-Ray Flashes \citep[XRFs,][]{2001grba.conf...16H}. 
Given the similarity of durations between XRFs, X-ray rich bursts and GRBs
\citep{2001grba.conf...16H,2003A&A...400.1021B}, the continuum of spectral
properties observed \citep*{2003astro.ph.12634L}, from classic GRBs to X-ray
rich bursts to XRFs and their cosmological origin
\citep{2003ApJ...599..957B,2003astro.ph.11050S}, it seems probable that XRFs and GRBs have very
similar origins. XRFs have been defined by a larger X-ray than $\gamma$-ray
fluence in the burst \citep[$S_{\rm X}/S_\gamma>1$,][]{2003astro.ph.12634L}; their
distinguishing characteristic is a shift in the peak energy ($E_{\rm peak}$) of the burst
from GRBs \citep[typically $\sim200$\,keV,][]{2000ApJS..126...19P} to XRFs
\citep[$\lesssim50$\,keV,][]{2000ApJS..126...19P,2003astro.ph.12634L}.
While GRBs and XRFs are located at cosmological distances, few have been
located at redshifts $<0.3$.  They are, GRB\,030329 at $z=0.1685$
\citep[associated with SN2003dh,][]{2003Natur.423..847H,2003ApJ...591L..17S},
XRF\,020903 with its probable host galaxy at $z=0.251$ \citep{2003astro.ph.11050S}, and
GRB\,980425 probably associated with SN1998bw at $z=0.0085$
\citep{1998Natur.395..670G}.  Of these, only the redshift for GRB\,030329 was
found directly from the afterglow and it has characteristics that are
still a matter of some debate; XRF\,020903 had a very low peak spectral
energy and a low luminosity \citep{2003astro.ph..9455S,2003astro.ph.11050S},
and GRB\,980425 had an extraordinarily low luminosity
\citep{1998Natur.395..663K}.  On the basis of GRB\,980425, it was suggested
that a variety of GRB with very low luminosity, detectable at low redshift,
form a distinct class of objects from classical, high-redshift GRBs
\citep{1998Natur.395..663K}.  It has also been suggested that low luminosity
in a burst is related to a steeply declining afterglow
\citep*[`f-GRBs',][]{2003ApJ...594..674B}, or to low $E_{\rm peak}$
\citep{2002A&A...390...81A,2003A&A...400.1021B,2003astro.ph..9455S}.

%
In this paper we report on \emph{XMM-Newton} observations of
\thisgrb,\footnote{We refer to this burst as \thisgrb\ to avoid confusion, though we show it to be an XRF in \sect\ref{xrf}.}
begun six hours after the burst, as well as a second,
serendipitous observation made on 6~Dec.\ 2003.  During the first observation an echo of the prompt
emission, reflected off dust in our Galaxy, was detected as an expanding
ring centred on the afterglow.  This was reported by
\citet{2003astro.ph.12603V} and that remarkable detection (the first time
that we are aware that prompt X-rays have been detected at energies below
2\,keV) has allowed us to infer properties of the burst itself in soft
X-rays, however the dust echo is not directly considered here.

A cosmology with $\Omega_{\rm m} = 0.3$, $\Omega_\Lambda = 0.7$ and
H$_0=75$\,km\,s$^{-1}$\,Mpc$^{-1}$ is assumed throughout. Error ranges
quoted are 90\% confidence intervals, unless stated otherwise.


%
%
\section{Observations and data reduction\label{observations}}

An initial 56\,ks exposure, centred on the error-circle provided by the
Integral Burst Alert System \citep{2003A&A...411L.291M,2003GCN..2459....1G},
was made with \emph{XMM-Newton}'s EPIC cameras. One X-ray source close to
the centre of the circle was observed to fade \citep{2003GCN..2474....1R}. 
The astrometry of the field was corrected by matching X-ray sources to stars
from the USNO-A2 catalogue using the SAS task \texttt{eposcorr}
\citep{Tedds:2000}.  The position of the fading X-ray source, (J2000)
R.\,A.\ = $08^h02^m30.190^s$, Dec.\ =
$-39$\degr51\arcmin04.05\arcsec,\ with a $1\sigma$ error radius of
0.7\arcsec\ \citep{2003GCN..2490....1T} is consistent with the position of a
fading radio source (\citealp{2003GCN..2473....1F}; \citealp*{2003GCN..2483....1S}).
A star-forming galaxy at redshift $z=0.105$
\citep{2003GCN..2470....1H,2003GCN..2481....1B,2004astro.ph..2085P} was also
detected within 0.5\arcsec\ of this position at near infrared wavelengths. 
No other fading sources are detected with EPIC in the IBAS error-circle.  In
light of this and the fact that the dust echo of a very bright transient
source is detected centred on the fading source \citep{2003astro.ph.12603V},
the X-ray source at this position must be the GRB afterglow.

The first \emph{XMM-Newton} observation was carried out under low background
conditions (the background represents $\sim10$\% of the extracted counts) and the data were not filtered for this reason
beyond standard processing.  Spectral data from the second observation were
filtered due to high background in the middle and end of the observation. 
The `thin' and `medium' filters were used in the first observation with the
pn and MOS cameras respectively. The `thick' filters were used with each
camera during the second observation, in which the principle target was
$\zeta$~Puppis, and the afterglow was off-axis by $\sim8\arcmin$. Data from
the EPIC-MOS and pn cameras are consistent within cross-calibration
uncertainties. Because of the
extreme lack of counts at low energies, most of the counts at energies below
0.8\,keV are redistributed from higher energies; low-level uncertainty in
this redistribution, usually negligible in spectra with only moderate
soft absorption, has caused us to exclude data below 0.8\,keV from the
spectral analysis.
The data reduction followed a standard procedure similar to that outlined in
\citet{2002A&A...395L..41W} except that the data were processed and reduced
with the \emph{XMM-Newton} SAS v.\,5.4.1.  A spectral binning using a
minimum of 20 counts/bin was used.  Two regions on the detector were
used to obtain different background spectra, avoiding obvious sources and the dust-echo. No significant difference was
observed using either background region.  To calibrate the absolute fluxes
for the lightcurves, the mean flux for the observation, derived from the
best-fit model to the joint fit of the EPIC spectra, was used to scale the
count-rate lightcurves, using the response matrices ancillary response
vectors generated by the SAS tasks \texttt{rmfgen} and
\texttt{arfgen}, which take into account, among other things, the off-axis
angle of the source and the filter used.  Careful analysis of
the flux calibration required the use of ancillary response vectors newer
than those available in the current version of the SAS; though usually
insignificant for on-axis sources, this does affect sources far from the
optical axis considerably (Saxton, priv.\ comm.).

The first observation was divided into five 11\,ks segments
to examine spectral evolution.

%
%

\section{Results\label{results}}
\begin{figure}
 \includegraphics[bb=58 79 554 740,angle=-90,width=\columnwidth,clip=]{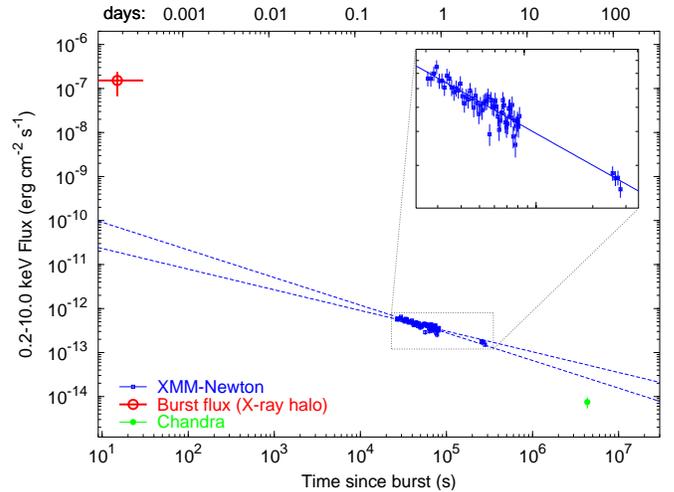}
 \caption{
          X-ray lightcurve (based on the mean EPIC 0.2--10.0\,keV absorbed
          fluxes) for the afterglow of \protect\thisgrb\ fit with a single
          power-law model.  The $3\sigma$ limits of the fit are plotted on
          the main figure (dashed line) while the best fit is shown on the
          inset. The temporal slope derived from the \emph{XMM-Newton}
          observations is much too flat ($-0.55\pm0.05$) to fit the prompt
          flux (open circle) as derived from the observed dust echo.  A
          slope $\lesssim-1.7$ is required to connect the prompt emission
          with the afterglow flux.  The late time flux detected by
          \emph{Chandra} \protect\citep{2004GCN..2522....1F} is shown for
          completeness and implies a steepening in the lightcurve after
          about three days.
         }
 \label{fig:lightcurve}
\end{figure}

The afterglow decays \citep{2003GCN..2474....1R} during
the first observation (Fig.~\ref{fig:lightcurve}), and is well fit
($\chi^2 = 44.9$ for 54 degrees of freedom) by a power-law with index $-0.54\pm0.09$,
a decay slower than reported for any previous X-ray afterglow except
GRB\,980425 \citep{2000ApJ...536..778P}.
Continuing the power-law decay observed in the first
observation fits the second observation well ($\chi^2 = 46.6$ for both
datasets).  The fit to both datasets, a power-law with index $-0.55\pm0.05$,
is not improved significantly by the addition of a break to the single
power-law model ($\chi^2_\nu = 45.6/56$ compared to $\chi^2_\nu = 46.3/58$, $f$-test probability = 0.65).

We \citep{2003astro.ph.12603V} have discovered the first dust echo
observed from a GRB and using it, have derived the 0.2--10.0\,keV
absorbed flux of the prompt emission
($1.5\pm0.8\times10^{-7}$\,erg\,cm$^{-2}$\,s$^{-1}$).  The drop in flux to
$5\times10^{-13}$\,erg\,cm$^{-2}$\,s$^{-1}$ observed in the afterglow at the
beginning of the \emph{XMM-Newton} observation requires a slope
$\sim-1.7$ and is not consistent with the decay slope of the best-fitting
power-law.  There must therefore have been a discontinuous flux
change of orders of magnitude between the prompt and afterglow phases in the
soft X-ray band \citep[a characteristic somewhat atypical of bursts observed
by \emph{BeppoSAX},][]{2000ApJS..127...59F}; or there was a faster decay
rate in the afterglow phase than is usual in early X-ray afterglows
\citep[typical early X-ray afterglow slopes are $\sim-1$ at $t\lesssim1$\,day
e.g.,][]{2003ApJ...597.1010B,2003A&A...409..983T,2002A&A...393L...1W,2002A&A...395L..41W}
which then slowed markedly in the afterglow phase.  This lightcurve is reminiscent of the optical
lightcurves of GRBs\,990123 and 021211, where a rapid decline was observed
initially ($\lesssim0.1$\,day), then a slower decay and finally a
resteepening
\citep{1999ApJ...519L..13F,1999Sci...283.2069C,2003ApJ...586L...9L}.


The complete spectra of the afterglow from the first and second observations
and the five 11\,ks spectra from the first observation can be fit with a
power-law with absorption in excess of the Galactic value
\citep[$5.9\times10^{21}$\,cm$^{-2}$,][]{1990ARA&A..28..215D}. 
The best-fit power-law photon indices ($N(E)\propto
E^{-\Gamma}$) are $1.90\pm0.05$ and $1.7\pm0.2$ for the first and second observations
respectively when the absorption was fixed to that found for the combined
fit to the dust echo and the afterglow spectra
\cite[$8.8\times10^{21}$\,cm$^{-2}$,][]{2003astro.ph.12603V}.  The
absorption is consistently $\sim50\%$ above the Galactic column
density, but this may be due to variations in the local column density on
scales $\lesssim1\degr$ \citep{1990ARA&A..28..215D}.  The observed,
time-averaged fluxes are $4.2\pm0.1\times10^{-13}$\,erg\,cm$^{-2}$\,s$^{-1}$
and $1.8\pm0.1\times10^{-13}$\,erg\,cm$^{-2}$\,s$^{-1}$.
\begin{figure}
 \includegraphics[width=1.0\columnwidth,clip]{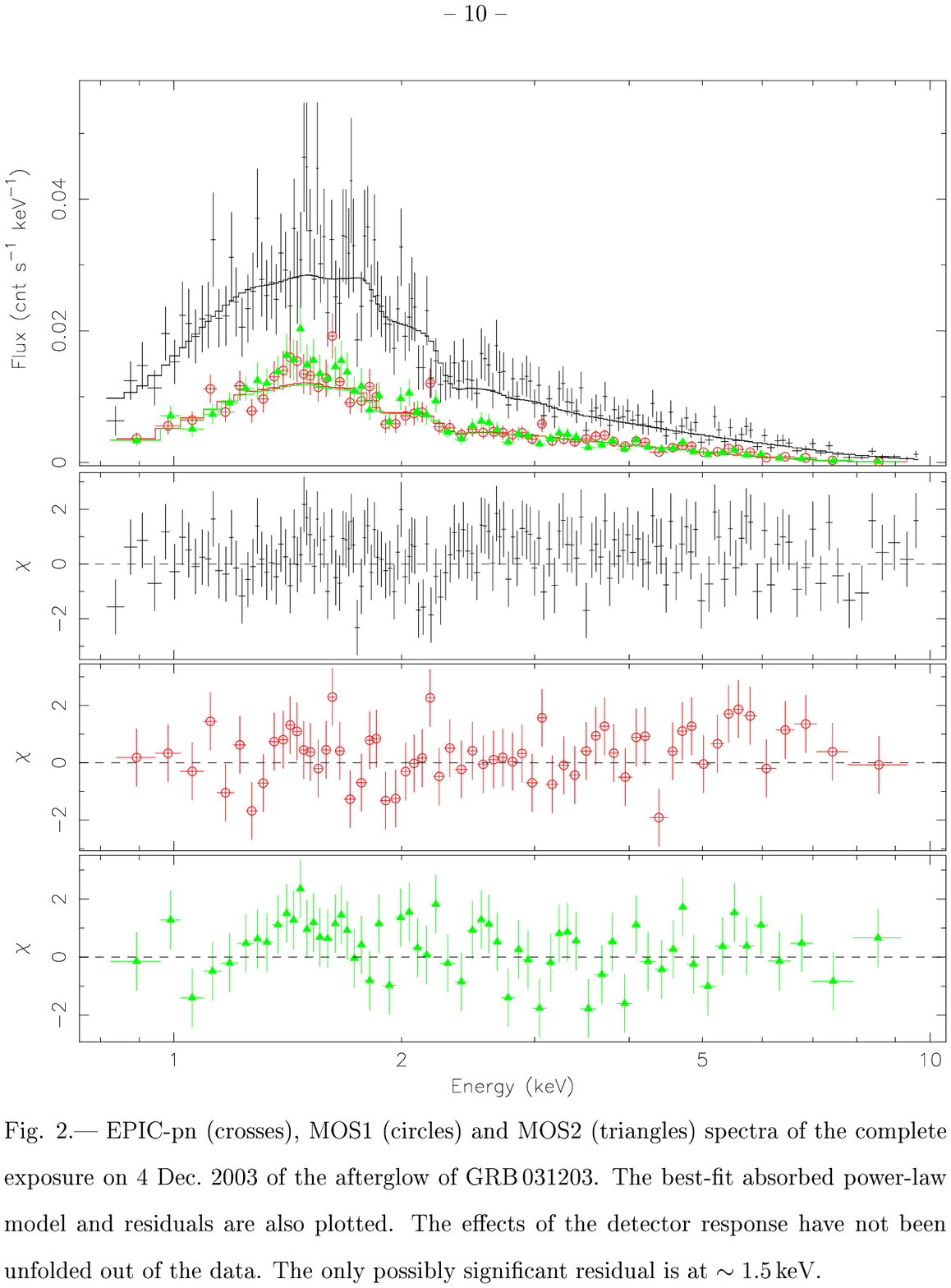}
 \caption{EPIC-pn (crosses), MOS1 (circles) and MOS2 (triangles) spectra
          of the complete exposure on 4 Dec.\ 2003 of the afterglow of
          \thisgrb.  The best-fit absorbed power-law model and residuals
          are also plotted.  The effects of the detector response have not
          been unfolded out of the data. 
         }
 \label{untouched_spectrum}
\end{figure}

%
%
\section{A low luminosity burst at $z=0.105$\label{assoc}\label{discussion}}

No optical afterglow was detected for this burst; the upper limit on the
flux of the transient based on an extrapolation of the X-ray
power-law at $\sim0.5$\,day ($\Gamma=1.9$), is $I>21.9$\,mag, including a
reddening correction of A$_I = 2.0$\,mag.  Given the brightness of the
underlying galaxy \citep[$I=19.1$\,mag,][]{2003GCN..2486....1B} a source
with observed $I>21.9$\,mag seems unlikely to be detected. 
The non-detection of the optical afterglow makes the
association of the GRB with the star-forming galaxy at $z=0.105$
uncertain since we do not have a redshift associated with the afterglow. The
X-ray observations strengthen this association. The X-ray position is the
best localisation available and places the GRB within
$\sim1$\,kpc of the centre of the galaxy; the probability (based on the
I-band galaxy counts of \citet{1998ApJ...506...33P} and neglecting
gravitational lensing and extragalactic absorption) of the centre of a
galaxy at least this bright appearing coincident with the \emph{XMM-Newton}
position by chance is low, $\sim7\times10^{-5}$.  The probability of the
centre of \emph{any} galaxy being detected (assuming a limiting I-band
magnitude of 23) within 0.7\arcsec radius is only 0.004.

At a luminosity distance of 453\,Mpc, \thisgrb\ is the closest GRB/XRF with
known redshift after GRB\,980425.  This distance implies that the
$\gamma$-ray burst energy and afterglow luminosity were very low compared to
cosmological GRBs \citep{2003GCN..2483....1S}.

\citet{2003GCN..2460....1M} quote a peak 20--200\,keV photon flux for
\thisgrb\ of 1.2\,cm$^{-2}$\,s$^{-1}$ and a duration of $\sim30$\,s. 
For the following calculations we use an estimate of the mean flux of one
third the peak flux \citep{2000ApJ...538..165H} to derive a photon fluence
of 12\,cm$^{-2}$.  We also use a power-law index
$\Gamma=2.5\pm0.3$ for the burst, justified below (\sect\ref{xrf}).

The equivalent isotropic $\gamma$-ray energy (20--2000\,keV rest-frame) is
only $3\times10^{49}$\,erg, well below the total $\gamma$-ray energy in the
general case \citep[$\sim10^{51}$\,erg,][]{2003ApJ...594..674B}, but a few
times greater than XRF\,020903
\citep{2003astro.ph..9455S} and $\sim30$ times that of
GRB\,980425 \citep{2000ApJ...536..778P}. The peak radio luminosity is often
used as a surrogate for the total burst energy; \thisgrb's radio luminosity
after 4\,days \citep{2003GCN..2473....1F} is very low, but within a factor
of three of GRB\,980425/SN1998bw at the same time
\citep{1998Natur.395..663K,2003astro.ph.11050S}, indicating that it may have an unusually low
peak radio luminosity. The X-ray luminosity for \thisgrb\ is also low
\citet{2003GCN..2483....1S}.  Even the isotropic equivalent luminosity based
on the observed 2--10\,keV flux is
$9\times10^{42}$\,erg\,s$^{-1}$ at 10\,hours, more than three
orders of magnitude below the typical value
($\sim5\times10^{46}$\,erg\,s$^{-1}$) \citep*{2003ApJ...590..379B}, though a
factor of $\sim100$ above GRB\,980425 \citep{2000ApJ...536..778P}.

\section{\thisgrb\ as an X-ray flash\label{xrf}}
The photon power-law index from the 0.2--10.0\,keV photon fluence (estimated
from the dust echo) to the 20--200\,keV photon fluence is
$\Gamma=2.5\pm0.3$.  Furthermore, a power-law fit to the dust echo,
corrected for the effects of the dust scattering, yields an intrinsic
$\Gamma=2.2\pm0.3$.  A single power-law (with $\Gamma\gtrsim2$) from 1 to
200\,keV is consistent with these numbers: in this case the $E_{\rm peak}$ of
the burst must be below 1\,keV, a remarkably low $E_{\rm peak}$ even for an
XRF. They are also consistent with a power-law index that steepens between
the soft (0.2--10.0\,keV) and hard (20--200\,keV) X-ray bands, with 
$E_{\rm peak}$ probably below the 20--200\,keV band.  It is likely that the 
$E_{\rm peak}$ of \thisgrb\ was $\lesssim20$\,keV.  \citet{2003GCN..2485....1D}
predicted $E_{\rm peak}$ for this burst close to 20\,keV based on the
cannon-ball model; we cannot exclude this possibility.  In any case,
\thisgrb\ is an XRF; the ratio used to define XRFs,
$\log({S_{\rm X_{\rm2-30\,keV}}/S_{\gamma_{\rm30-400\,keV}}})>0$, is
$0.6\pm0.3$ in this case, an extreme value \citep[compare
XRF\,020903 with a ratio of 0.7,][]{2003astro.ph..9455S}, assuming a single
power-law in the 2--400\,keV band. 

We assume above that the fluence observed in the X-ray halo is due
entirely to the burst event.  Some fraction will, of course, be
contributed by the afterglow, since the time of the observed X-ray fluence
is only constrained to $\sim1$ hour of the burst.  To quantify this, we
derive the maximum afterglow contribution to the fluence (where the
afterglow flux starts at the mean burst flux and decays without a break to
the flux observed at the start of the \emph{XMM-Newton} observations).  This
depends on the mean flux of the burst and therefore on the burst duration. 
The afterglow contributes $<60$\% of the total fluence when the burst
duration is 30\,s.  Even at 60\% \thisgrb\ is still an XRF, with
$\log(S_{\rm X}/S_\gamma)\sim0.3$, within the uncertainty mentioned above. 
The contribution of the afterglow to the observed soft X-ray fluence cannot
influence the classification of \thisgrb\ as an XRF. In order for \thisgrb\
not to be an XRF, a very large fraction of the fluence would have to be
significantly earlier or later than the hard X-ray burst and (somewhat
arbitrarily) renamed a ``precursor'' or a ``rebrightening'', with a total
energy comparable to the burst itself.

\emph{INTEGRAL}'s SPI-ACS detected \thisgrb; 
the count-rate is not inconsistent with that expected from the
fluence and spectral slope mentioned above \citep{2003astro.ph..2139V}.  If
the ratio of peak count-rate to mean count-rate \citep{2004astro.ph..2085P}
in the IBIS band (20-200\,keV) is the same as in SPI-ACS band ($>100$\,keV),
then the total 20-200\,keV fluence estimated above will be smaller by a
factor 3.2, making \thisgrb\ the most extreme XRF ever observed, with
$\log({S_{\rm X_{\rm2-30\,keV}}/S_{\gamma_{\rm30-400\,keV}}})\sim1$.
An analysis of all the \emph{INTEGRAL} data will be interesting since
the quoted peak flux for the burst
($1.3\times10^{-7}$\,erg\,cm$^{-2}$\,s$^{-1}$ in the 20-200\,keV band)
implies a much harder spectrum during the peak ($\Gamma\sim1.3$),
indicating strong spectral evolution.  This has no impact on
our calculations above since they are based on the total fluences
in the burst, derived from the count-rates only, not the peak flux.

A tantalising result we can compute is an approximate upper limit on the
pseudo-z \citep[$\hat{z}$,][]{2003A&A...407L...1A} using $E_{\rm peak}$
$<20$\,keV, $\Gamma>1.9$ and $t_{90} = 30$\,s, for which we find
$\hat{z}\lesssim0.15$.

\section{Discussion}
\citet{2002A&A...390...81A} have found a relation between the equivalent
isotropic $\gamma$-ray total energy ($E_{\rm iso}$) and $E_{\rm peak}$ in
GRBs.  Though only a single burst, the low luminosity XRF\,020903 has extended this relation to very
low luminosities and peak energies \citep{2003astro.ph..9455S}.  Using the
best-fit values for this relation quoted by \citet{2003astro.ph..9455S}, and
the equivalent isotropic luminosity in the 1\,keV--10\,MeV band derived here
($1.5\times10^{50}$\,erg), the $E_{\rm peak}$ of \thisgrb\ should lie close to
$\sim10$\,keV; this is consistent with the $E_{\rm peak}$ suggested above and
further confirms the correlation between peak energies and $E_{\rm iso}$.

One implication of this correlation is clear; XRFs have low $E_{\rm iso}$ as
we see here. Furthermore, if the total energy in $\gamma$-rays is nearly
constant at $\sim10^{51}$\,erg
\citep{2001ApJ...562L..55F,2003ApJ...594..674B}, it implies that the opening
angle in XRFs is extremely wide.  However, this becomes problematic if there
is a significant population of XRFs with peak energies below $\sim20$\,keV,
since the \citet{2002A&A...390...81A} relation implies $E_{\rm iso}$ for
these bursts is less than the total energy in
$\gamma$-rays inferred by \citet{2001ApJ...562L..55F}.  
Only one XRF so far had clearly indicated this paradox; XRF\,020903,
mentioned above.  But now \thisgrb\ similarly shows a very low $E_{\rm
iso}$.  (We note that it has been suggested that XRF\,030723 also has a very
low $E_{\rm iso}$ \citep{2003astro.ph.10414L}, but the redshift in that case
is highly uncertain, \citeauthor{2004astro.ph..2240F} \citeyear{2004astro.ph..2240F}).  These results hint that a significant population of
(so far, rare) low luminosity bursts may indeed exist as required in some
models of off-axis GRBs \citep[e.g.][]{2004astro.ph..1142Y}.

It has been suggested \citep{2003ApJ...594..674B} that a class of
underluminous GRBs exist with fast decay rates (f-GRBs) at early times
($t\lesssim0.5$\,days).  We cannot exclude \thisgrb\ from such a class. 
Both its luminosity and decay rates are extraordinary, the luminosity is
very low and the decay rate unusually slow, but there must be a fast decay in
the afterglow before the
\emph{XMM-Newton} observation at $<0.25$\,days, if the prompt and afterglow
fluxes decay smoothly and there is no abrupt drop in the flux between prompt
and afterglow phases.  GRB\,980425, possibly similar in terms of its low
luminosity, is also interesting as a comparison object in terms of its X-ray
decay structure. GRB\,980425 also showed a fast decay from the prompt X-ray
flux (power-law index, $\alpha\sim-1.5$) and a very slow decay in X-rays
from 1 to 100\,days ($\alpha\sim-0.2$). 
In XRF\,030723 a very slow decay was observed at optical wavelengths at
early times \citep{2004astro.ph..2240F}: perhaps very slow
early decay rates are related to the apparent luminosity and $E_{\rm peak}$
of the burst lending some support to the idea that these bursts may be
viewed off-axis \citep{2002ApJ...570L..61G,2004astro.ph..1142Y}. 

\acknowledgments
We thank R.~Saxton for help with the EPIC flux calibration
and the \emph{XMM-Newton} project scientist and SOC staff for continuing
rapid response GRB observations, recognising the
operational load these place on the project.  We acknowledge benefits from
collaboration within the EU FP5 Research Training Network, `Gamma-Ray
Bursts: An Enigma and a Tool'. This work was also supported by the Danish
Natural Science Research Council (SNF).


\end{document}